\documentclass[aps,prb,twocolumn,superscriptaddress,showpacs,floatfix]{revtex4-2}
\usepackage{comment}
\usepackage{physics}
\usepackage{amsmath,amsthm,amssymb,amsfonts, color, comment, graphicx}
\usepackage{xcolor}
\usepackage{float}
\usepackage{graphicx}
\usepackage{dcolumn}
\usepackage{bm}
\UseRawInputEncoding 
\usepackage{dsfont}
\usepackage{hyperref}
\usepackage[normalem]{ulem}

\newcommand{\be}{\begin{equation}}
\newcommand{\ee}{\end{equation}}
\newcommand{\bk}{{{\bf{k}}}}

\newcommand{\br}{{\bf{r}}}

\newcommand{\beal}{\begin{align}}
\newcommand{\eeal}{\end{align}}
\newcommand{\ra}{\rangle}
\newcommand{\la}{\langle}

\newcommand{\dg}{{\dagger}}
\newcommand{\pdg}{{\phantom\dagger}}

\renewcommand{\vec}[1]{\mathbf{#1}}

\usepackage{graphicx}
\usepackage[caption=false]{subfig}

\newcommand{\contribute}{$^{\phi}$}

\footnotetext{\contribute~These authors contributed equally to this work}

\begin{document}

\preprint{APS/123-QED}
\title{Pseudo-chiral phonons from octupolar magnetic order}
\author{Ruairidh Sutcliffe {\contribute}}
\email{ruairidh.sutcliffe@mail.utoronto.ca}
\affiliation{Department of Physics, University of Toronto, 60 St. George Street, Toronto, ON, M5S 1A7 Canada}
\author{Kathleen Hart {\contribute}}
\email{kf.hart@mail.utoronto.ca}
\affiliation{Department of Physics, University of Toronto, 60 St. George Street, Toronto, ON, M5S 1A7 Canada}
\author{Swati Chaudhary {\contribute}}
\email{swatichaudhary@issp.u-tokyo.ac.jp}
\affiliation{The Institute for Solid State Physics, The University of Tokyo, Kashiwa, Chiba 277-8581, Japan}
\author{Arun Paramekanti}
\email{arun.paramekanti@utoronto.ca}
\affiliation{Department of Physics, University of Toronto, 60 St. George Street, Toronto, ON, M5S 1A7 Canada}


\begin{abstract}
Motivated by the recent discovery of anomalously large magnetic response of chiral phonons in dipolar magnets, we introduce
the concept of pseudo-chiral phonons which are shown to emerge in multipolar magnets. We consider
Raman active quantum phonons, such as doublet $E_g$ phonons $(d_{x^2-y^2},d_{3z^2-r^2})$ in cubic crystals, which feature a 
symmetry-allowed \emph{linear} coupling to local quadrupolar moments.
We show that distinct multipolar orders can imprint distinct patterns of degeneracy breaking for phonons, with
quadrupolar orders favoring non-chiral eigenmodes and time-reversal breaking octupolar orders 
favoring unconventional pseudo-chiral phonons.
We compute the temperature dependent phonon matrix Green function
using a path integral approach where `fast' phonon modes sense `slow' pseudospin fluctuations over a thermal background sampled
using Monte Carlo simulations.
We propose helicity-resolved Raman spectroscopy of these pseudo-chiral phonons as a probe of hidden octupolar order in quantum materials 
such as Ba$_2$CaOsO$_6$ and PrV$_2$Al$_{20}$.
\end{abstract}

\maketitle

Phonon chirality has emerged as an important concept in quantum materials, highlighting the strong interplay between
phononic and other degrees of freedom such as spin, charge, and orbitals \cite{Chen2019,Wang2024,zhang2025chirality,zhang2025new,yang2025catalogue,juraschek2025chiral}. Chiral phonons are characterized by a nonzero angular momentum and are commonly associated with circularly polarized lattice vibrations~\cite{Zhang2015PRL,Zhu2018,ueda2023chiral,ishito2023truly,ueda2025chiral,ohe2024chirality,minakova2025direct,Chen2019_entangle}, though anomalous
rotationless chiral phonons have also been identified~\cite{chaudhary2025anomalous}.
These phonons have attracted a significant attention due to their strong magnetophononic~\cite{Ren2021,hernandez2023observation,ZhangPRL2023,cheng2020large,Chen2025chiral,Baydin2022,chaudhary2024giant,lujan2024spin,mustafa2025,BowenPRL2024,chen2025geometric,Bistoni2021,RenPRX2024,wang2025ab,Che2025PRL,Yang2025PRL,boniniprl2023,wu2025magnetic,wu2023fluctuation,cui2023chirality,wu2023fluctuation,Ren2024PRL,shabala2025axial} and phono-magnetic responses~\cite{juraschek2020giant,nova2017effective,luo2023large,basini2024terahertz,Davies2024,romao2024light,Xiong2022,Shabala2024,juraschek2017dynamical,geilhufe2021dynamically,basini2024terahertz,hart2024phonon,Guohuan2022,li2021,cui2023chirality,ma2023chiral,YaoPRB2025,nishimura2025theory,kim2023chiral}.

 Chiral phonons inherit signatures of time-reversal symmetry (TRS) breaking in the electronic, orbital, or spin sectors. These signatures manifest as chirality-dependent energy splitting and have been observed  across a broad range of materials, including both magnetic~\cite{lujan2024spin,Che2025PRL,Yang2025PRL} and non-magnetic systems~\cite{hernandez2023observation,cheng2020large}. In non-magnetic materials, chirality and mode splitting often arise from underlying nontrivial electronic band topology~\cite{Hu2021,Ren2021}. In contrast, in magnetic materials, linearly polarized doubly degenerate phonon modes can split into two finite angular momentum-carrying modes either under an external magnetic field~\cite{lujan2024spin,schaack:1976,schaack:1977} or due to intrinsic dipolar magnetic order~\cite{Che2025PRL,Yang2025PRL,boniniprl2023,wu2025magnetic}. Experimentally, these phonons can be detected via helicity-resolved Raman spectroscopy where the two chiralities appear in opposite cross-circular polarization channels~\cite{ishito2023truly,lujan2024spin,Che2025PRL,wu2025magnetic,mustafa2025}. These studies indicate that phonon chirality can serve as a distinct 
 probe of non-trivial band topology or dipolar magnetic orders.

\begin{figure}[t]
\includegraphics[width=0.45\textwidth]{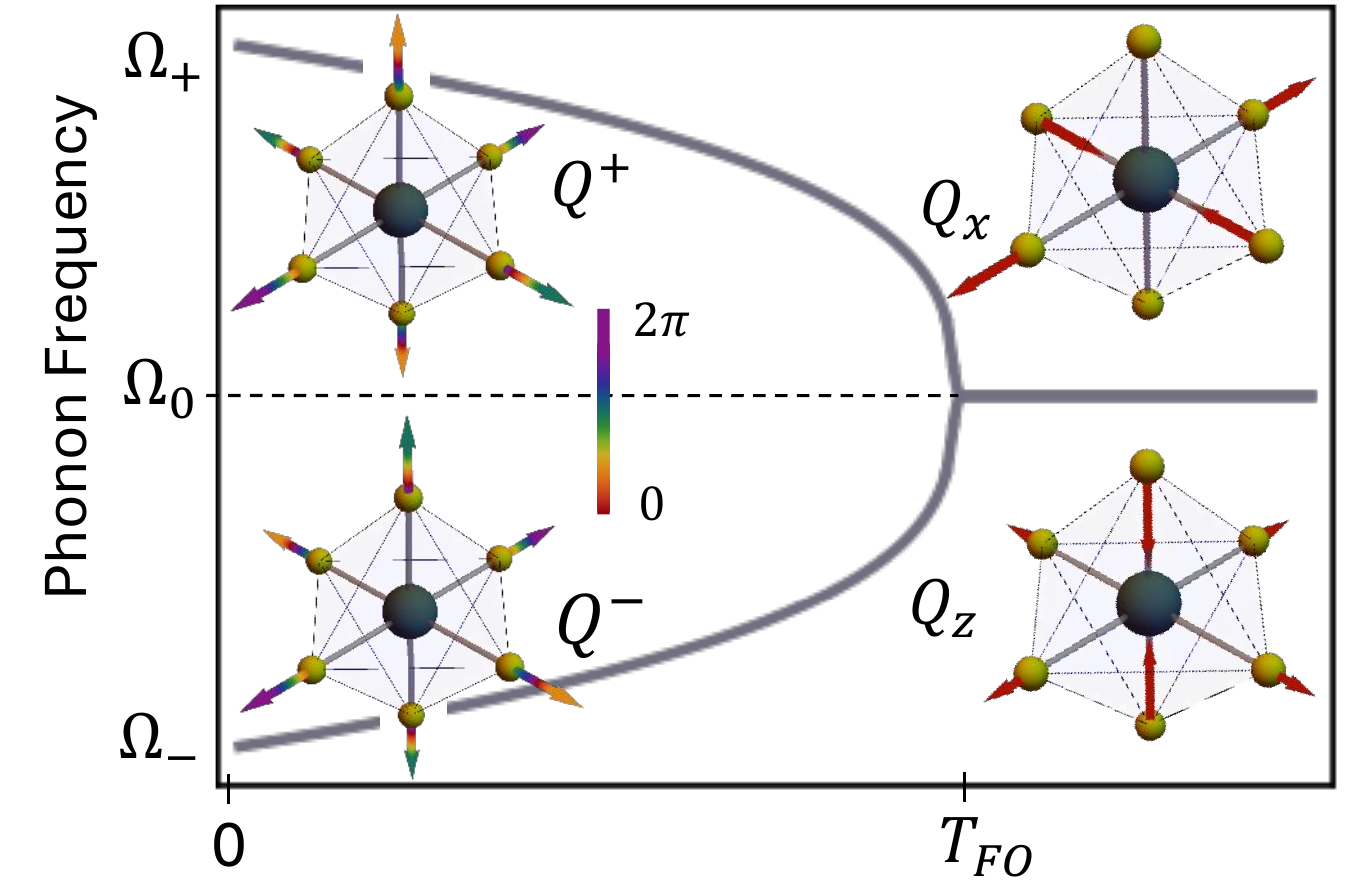}
\caption{Schematic figure showing degenerate ${\cal E}_g$ phonon modes $(Q_x,Q_z) \equiv (d_{x^2-y^2},d_{3z^2-r^2})$ with frequency $\Omega_0$
of an octahedral cage around a central magnetic ion hosting a non-Kramers doublet with quadrupolar and octupolar moments. In a ferrooctupolar 
(FO) phase ($T < T_{\rm FO}$)
these split into pseudo-chiral eigenmodes $Q^{\pm} \equiv Q_x \pm i Q_z$ with frequencies $\Omega_\pm$ and the indicated phases for
the displacements. The sign of the splitting depends on the sign of the Ising FO order. See Supplemental Material \cite{suppmat} for
a video of the $Q^\pm$ modes.}
\label{fig:specFuncs}
\end{figure}

 Beyond conventional magnetic dipole orders, a variety of quantum materials exhibit magnetic ordering which is 
 described in terms of higher-order or 
 higher-rank multipoles~\cite{Kuramoto2008,Kuramoto2009}. Familiar examples include spin nematic order in a spin-$1$
 magnet which hosts rank-2 quadrupolar order \cite{tsunetsugu2006nematic,Penc2011}, 
 and $d$-wave altermagnetic order \cite{Smejkal2022, Bhowal2024a,jungwirth2024,jaeschkeubiergo2025atomicaltermagnetism} 
 which has a rank-3 octupolar magnetization density
 in the unit cell \cite{Bhowal2024a,jaeschkeubiergo2025atomicaltermagnetism}.
In this work, we ask how does such multipolar ordering affect the phononic sector? Can it induce phonon chirality, creating 
non-degenerate phonon modes with nonzero angular momentum, similar to dipolar systems? 
Could the multipolar imprint on phonon spectra and chirality offer a new route for detecting these `hidden orders' which evade 
conventional probes?

In this Letter, we explore these questions for localized multipole moments which are ubiquitous 
in quantum materials with strong spin-orbit coupling. These include actinides NpO$_2$ \cite{Santini2006,Santini2009} 
and $\rm{URu_2Si_2}$ \cite{chandra2002hidden,cricchio2009itinerant,Haule2009,mydosh2011colloquium},
lanthanides such as
$\rm{CeB_6}$ \cite{Shiina1997,Ye2019b,Otsuki2024}, PrO$_2$ \cite{Khmelevskyi2024}, Pr$M_2$Al$_{20}$ ($M$=V, Ti) \cite{ShimuraPRB2015,matsumoto2016strong,ishitobi2021triple,lee2018landau,hattori2014antiferro}, spin-ice compounds
 Pr${_2}M_2$O$_7$ ($M$=Sn,Hf,Zr) and
Tb${_2}$Ti$_2$O$_7$ 
\cite{patri2020theory,gritsenko2020changes,princep2013crystal},
and transition metal oxides $\rm{Ba_2NaOsO_6} $ \cite{Chen2010,Lu2017},
$\rm{Ba_2MgReO_6} $ \cite{hirai2020detection,Soh2024,mosca2024interplay}, $\rm{Sr_2MgReO_6}$ \cite{fiore2025antiferro},
and {Ba$_2 M$OsO$_6$}
($M\!=$Cd,Mg,Ca)~\cite{Chen2011,voleti2021octupolar,paramekanti2020octupolar,maharaj2020octupolar,churchill2022competing}.
The local moments in these systems
encode higher-rank multipoles \cite{kuramoto2009multipole} such as 
 time-reversal even (${\cal T}$-even) rank-2 quadrupoles and rank-4 hexadecapoles
as well as time-reversal odd (${\cal T}$-odd) rank-3 octupoles and rank-5 trikontadipoles.
Intriguingly, a
subset of these multipolar operators, namely
$\mathcal{T}\rm{-even}$ quadrupolar moments, exhibit a generic {\it linear} coupling 
to Raman-active 
phonons. The broad message of our work
is that this unusual coupling enables multipolar orders to leave their imprint on the 
phonon dynamics. In particular, $\mathcal{T}\rm{-odd}$ octupolar order engenders
{\it pseudo-chiral} phonon eigenmodes,
enabling a distinct route to probing 
unconventional broken symmetries via Raman spectroscopy of phonon modes.
 
\noindent{\bf Pseudo-chiral phonons:} We begin by introducing the concept of pseudo-chiral 
phonons. Let us consider a cubic crystal with degenerate ${\cal E}_g$ phonon modes $(Q_x,Q_z) \equiv (d_{x^2-y^2},d_{3z^2-r^2})$.
Conventional symmetry breaking associated with lattice distortions can lift this degeneracy
while preserving time-reversal symmetry. For instance, a tetragonal distortion will lead to $Q_x$ and $Q_z$
becoming non-degenerate phonon eigenmodes. Remarkably, we can also construct time-reversal breaking phonon
eigenmodes $Q_x \pm i Q_z$. These phonon modes do not result in circular atomic motion and do not carry 
real angular momentum unlike chiral phonons in 
dipolar magnets \cite{wu2025magnetic,Che2025PRL}; indeed, the ordinary angular momentum
is quenched for ${\cal E}_g$ phonons. Instead, these $Q^{\pm}$ modes are 
characterized by the phase structure shown in Fig.~\ref{fig:specFuncs} which
endows them with a pseudo-angular momentum (PAM) along the $C_3$ axis of the cubic crystal,
$\hat n \parallel [111]$ (and equivalent cubic body diagonals).
This phase structure defines the phonon PAM  $l$ along $\hat{n}$ as
\begin{equation}
C_3^{\hat{n}}Q^{\pm}=e^{\pm i \frac{2\pi}{3} l}Q^{\pm}
\end{equation}
where $l$ can take values of $0,1,2$. 
With this definition, the two pseudo-chiral phonon modes $Q_x \pm i Q_z$
carry opposite PAM (see SM \cite{suppmat} for details and a video of the two
modes) and, hence would be visible 
in opposite cross-circular channels in Raman spectroscopy \cite{suppmat}. 
In the remainder of this Letter, we discuss
how pseudo-chiral phonons can emerge in the presence of octupolar magnetic order, and how they might
shed light on competing multipolar symmetry breaking orders.

\noindent{\bf Multipole-phonon coupling:}
As an illustrative example of how diverse multipolar orders can imprint on phonon dynamics, 
we focus here on competing quadrupolar and octupolar degrees of
freedom which emerge for Pr$^{3+}$ moments on the diamond lattice in
{Pr$M_2$Al$_{20}$} ($M$=Ti,V) \cite{sakai2011kondo,sakai2012superconductivity,ShimuraPRB2015,taniguchi2016nmr,Taniguchi2019,patri2019unveiling,Ye2023}, 
and Os$^{6+}$ or W$^{4+}$ moments on
face-centered cubic (fcc) lattices in {Ba${_2}M$OsO$_6$} ($M$=Ca,Mg) \cite{Chen2011,paramekanti2020octupolar,maharaj2020octupolar,khaliullin2021exchange,voleti2021octupolar,churchill2022competing,Cong2023,voleti2023probing}
and Cs$_2$WCl$_6$ \cite{Morgan2023,Pradhan2024,Li2024}.
The competing multipolar degrees of
freedom in these cubic materials are encoded in local pseudospin-1/2 doublets protected by crystalline point group symmetries,
so-called ``non-Kramers doublets''. The pseudospin operators can be labelled
$\boldsymbol{\tau} = (\tau_x, \tau_y, \tau_z)$. Here, $\tau_x \propto (J_x^2-J_y^2)$ and $\tau_z \propto (3 J_z^2-J(J+1))$ encode quadrupolar moments, 
while the Ising-like octupole moment is $\tau_y \propto \overline{J_x J_y J_z}$
with the overbar denoting operator symmetrization. Here, $J$ is the angular momentum of the ion, with $J\!=\!2$ for Os$^{6+}$ or W$^{4+}$
and $J\!=\! 4$ for Pr$^{3+}$, which is split by the local crystal field environent. 
We investigate below the signatures of quadrupolar and octupolar order on $E_g$ phonon modes $(Q_x,Q_z) \equiv 
(d_{x^2-y^2},d_{3z^2-r^2})$ 
, and show that both types of ordering lift the phonon degeneracy, but in qualitatively distinct ways.
For quadrupolar orders, the new phonon eigenmodes are real superpositions
of $Q_x,Q_z$ and do not carry PAM.
By contrast, for octupolar order, the new phonon eigenmodes are uniform pseudo-chiral superpositions 
$Q^\pm=Q_x\pm iQ_z$ which carry PAM.


For the face-centered cubic lattice, 
symmetries dictate the form of the pseudospin Hamiltonian as,
\begin{eqnarray}
\!\! H_{\rm sp} \!&=&\!\!\! \sum_{\langle \br,\br'\rangle}\!\!\! \left[ - J_0 \tau_{\br y} \tau_{\br' y}
\!+\! J_1 \left(\cos^2\!\phi_{\br\br'} \!+\!  \gamma  \sin^2\!\phi_{\br\br'} \right) \tau_{\br z} \tau_{\br' z} \right. \nonumber \\ 
&+& J_1 \left( 1-\gamma \right) \sin\phi_{\br\br'} \cos\phi_{\br\br'} \left( \tau_{\br x}\tau_{\br' z} + \tau_{\br z}\tau_{\br' x} \right) \nonumber \\
&+&  \left. J_1 \left(\sin^2\phi_{\br\br'} + \gamma \cos^2\phi_{\br\br'}  \right)\tau_{\br x} \tau_{\br' x} \right],
\label{eq:nk1}
\end{eqnarray}
where $\phi_{\br\br'} = \{ 0 , 2\pi/3 , 4\pi/3 \}$ correspond to nearest neighbors $\la\br\br'\ra$ in the $\{ XY,YZ,ZX \}$ planes \cite{paramekanti2020octupolar,voleti2021octupolar}.
For instance, the {Ba$_2 M$OsO$_6$} compounds were proposed to have a ferro-octupolar (FO) ground state $\langle \tau_y \rangle \neq 0$, 
with $J_0 \!\sim\! 1$\,meV,
$J_1/J_0 \!\approx\! 0.5$ and $\gamma \!\approx\! -0.4$ \cite{pourovskii2021ferro}. However, an alternative antiferro-quadrupolar (AFQ) 
order with staggered $\langle \tau_{x,z}\rangle \neq 0$ has been suggested
as a competing ground state with $J_1/J_0 \!\gtrsim\! 2$ \cite{churchill2022competing,hart2024phonon}.
Given theoretical uncertainties in precisely determining these microscopic exchange couplings, and the absence of 
smoking gun experiments to distinguish various multipolar orders, the ground state of many of these systems remains debated.

To explore how these pseudospin degrees of freedom imprint on phonons and lead to useful experimental distinctions
between these competing orders, we focus on the two degenerate ${\cal E}_g$ phonon modes 
$(d_{x^2\!-\!y^2},d_{3z^2\!-\! r^2})$ with mode creation operators $(b^\dagger_x,b^\dagger_z)$ respectively. Treating these 
as Einstein phonons, the phonon Hamiltonian
\begin{equation}
\label{eq:nk2}
    H_{\rm ph} = \Omega_0  \sum_{\br} \sum_{\ell=x,z} b^\dagger_{\br\ell} b^\pdg_{\br\ell}
\end{equation}
where $\Omega_0$ is the $E_g$ mode frequency (we set $\hbar=1$). 
On symmetry grounds, the ${\cal T}$-odd  octupole will not linearly 
couple to phonons, however the $(\tau_x, \tau_z)$ quadrupoles can couple linearly to the phonon modes 
via a symmetry allowed on-site term
\begin{equation}
\label{eq:nk3}
    H_{\rm sp-ph} = - \lambda  \sum_\br \sum_{\ell=x,z} (b^\pdg_{\br\ell} +b^\dg_{\br\ell}) \tau^\pdg_{\br\ell}
\end{equation}
where $\lambda$ is the strength of the spin-phonon coupling.
For all the materials in question,
the phonon energy scale is much larger than the pseudospin exchange scale; e.g., \textcolor{black}{in
{Ba$_2 M$OsO$_6$}, density functional theory studies yield
$\Omega_0 \sim 60$\,meV \cite{okamoto2025spin,voleti2023probing}} so $\Omega_0 \gg J_0,J_1$. The spin-phonon
coupling $\lambda$ can also be estimated from a density-functional theory (DFT) calculation;
in $\rm{Ba_2CaOsO_6}$, this yields $\lambda \approx 5$\,meV \cite{voleti2023probing,hart2024phonon}.
Below, we use a path integral approach to compute the matrix phonon Green function 
in the limit where
`fast' phonon modes sense the `slow' pseudospin quantum fluctuations over a thermal 
background sampled in Monte Carlo (MC) simulations.

\noindent{\bf Pseudospin-phonon fluctuations:}
Since the phonon modes have $\Omega_0\!\gg \! J_0,J_1$,  we can treat these `fast' phonon modes as sensing the
`slow' local pseudospins via the quadrupole-phonon coupling. The pseudospin interactions in Eq.~\ref{eq:nk1} lead to an effective
spin Hamiltonian  $H_{\rm sp} = -\sum_\br \boldsymbol{h}_\br \cdot \boldsymbol{\tau}_\br$ where 
$\boldsymbol{h}_\br$ is the local Weiss field sensed  by $\boldsymbol{\tau}_\br$ arising from its neighboring pseudospins. 
$\boldsymbol{h}_\br$ is nearly uniform in the ordered phase and is randomly distributed
in the high temperature paramagnet. For a given background $\boldsymbol{h}_\br$, we
incorporate leading order spin-wave fluctuations and use this to study the impact on phonon modes,
and average the result over the temperature-dependent distribution
$P(\{\boldsymbol{h}_\br\})$ which we extract using large scale MC simulations.

To study excitations, it is useful to recast the local spin Hamiltonian in terms of Holstein-Primakoff bosons. To focus on transverse spin fluctuations around
the local Weiss field direction, we define a local reference frame $\{\tau'_x, \tau'_y, \tau'_z\}$, via
a rotation ${\boldsymbol{\tau}} \!=\! R(\theta_\br,\phi_\br)\!\cdot\!\vec{\tau'}$ 
where $\theta_\br$ and $\phi_\br$ are the spherical polar and azimuthal angles respectively, chosen such that ${\tau}_y'$ points along 
$\boldsymbol{h}_\br$. In this frame, we define boson operators $\alpha_\br, \alpha_\br^\dagger$ such that 
$
    \tau'_{\br y} = 2 (S - \alpha^\dagger_\br \alpha_\br),\,
    \tau'_{\br x} = \sqrt{2 S} (\alpha_\br + \alpha^\dagger_\br),\,
    \tau'_{\br z} = i\sqrt{2 S} (\alpha_\br - \alpha^\dg_\br).
$


The bosonic Hamiltonian $H_{\rm sp} =- \sum_\br 2{h_\br} (S-\alpha^\dagger_\br\alpha_\br)$,
which yields a local spin-wave boson with energy $ 2{h_\br}$.
In this frame, the spin-phonon Hamiltonian reduces to
\begin{eqnarray}
    \!\!\!\!\!\!\!\! && H_{\rm sp-ph}= \! \lambda \sum_\br [(b_{\br x}\!+\! b^\dagger_{\br x})(\sqrt{2 S}\sin\phi_\br(\alpha_\br \!+\! \alpha^\dagger_\br) \nonumber \\
    \!\!&+& \! 2 \sin\theta_\br\cos\phi_\br(S \!-\! \alpha^\dagger_\br \alpha_\br)
    \!+\! i\sqrt{2 S}\cos\theta_\br\cos\phi_\br(\alpha^\pdg_\br \!-\! \alpha^\dg_\br)) \nonumber \\
    \!\!&+&\! (b^\pdg_{\br z}\!\!+\! b^\dg_{\br z})(2 \cos\theta_\br(S \!-\! \alpha^\dagger_\br \alpha_\br)
    \!+\! i\sqrt{2 S}\sin\theta_\br(\alpha^\pdg_\br \!-\! \alpha^\dg_\br))]
\end{eqnarray}
{This retains transverse quantum spin fluctuations, via Holstein-Primakoff bosons $\alpha_\br$,
around the classical background $\boldsymbol{h}_\br \equiv (h_\br,\theta_\br,\phi_\br)$ 
sampled by MC simulations.
).

\begin{figure*}[t]
\includegraphics[width=1.0\textwidth]{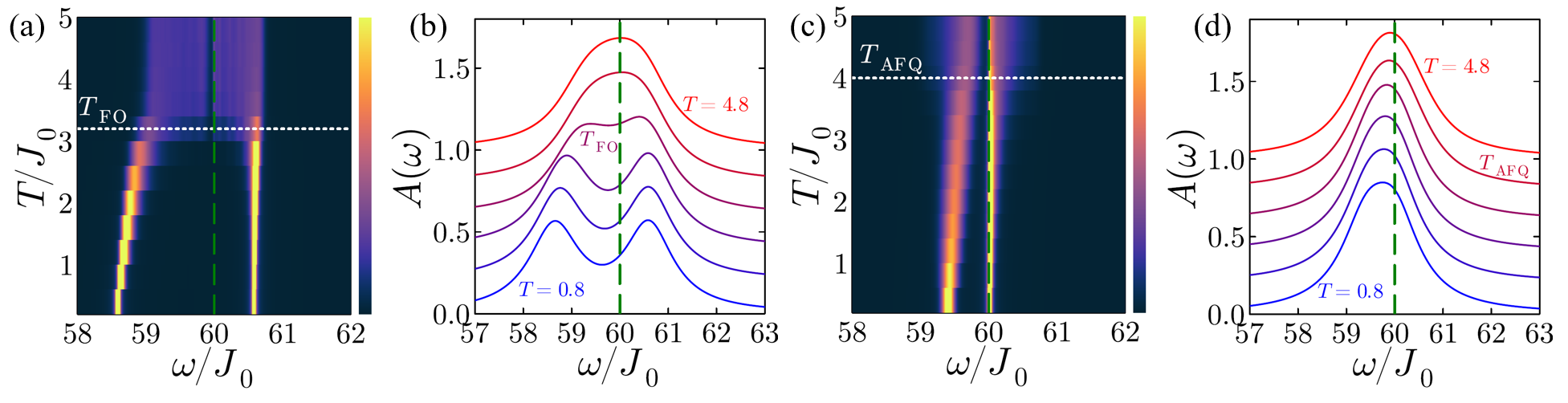}
\caption{\textcolor{black}{Phonon spectral function $A(\omega)$ of $E_g$ phonon modes $(d_{x^2-y^2},d_{3z^2-r^2})$
at equally spaced temperature values. (a),(b) With exchange parameter $J_1=1.3J_0$, which exhibits a low-temperature FO phase and a transition to a paramagnetic phase at $T_{\rm{FO}}\simeq3.2J_0$, a splitting of the phonon frequency is present. (a) There are two distinct peaks below $T_{\rm FO}$ in the undamped system. As the temperature increases, the peak heights decrease and the spectra broaden while preserving a `pseudogap' upto
$T^* \!\approx\! 1.2 T_{\rm FO}$. Eventually, $A(\omega)$ forms a broad spectrum centered at the degenerate natural phonon frequency $\Omega_0$ 
(green dotted line). 
(b) While a damping parameter of $2\pi\eta/\Omega_0 = 0.07$ broadens the spectra, 
the two peak structure is still distinguishable below $T_{\rm FO}$ (successive $T$ curves are vertically offset by $0.2$ for clarity). 
(c),(d) With exchange parameter $J_1=2.3J_0$ which exhibits a low-temperature quadrupolar ordered phase and a transition to a paramagnetic phase at $T_{\rm{AFQ}}\simeq4.0J_0$, there is a small splitting at low $T$ in the (c) undamped case that cannot be resolved with high damping of (d) $2\pi\eta/\Omega_0 = 0.07$. As temperature increases, the peak of $A(\omega)$ in (d) shifts to the degenerate natural phonon frequency $\Omega_0$.}}
\label{fig:specFuncs-local}
\end{figure*}

\noindent{\bf Effective phonon action:}
To second order in the spin-phonon coupling $\lambda$, the effective phonon action is
\begin{eqnarray}
    S^{\rm ph}_{\rm eff} \!=\! S^{\rm ph}_0 \!+\! \langle S_{\rm sp-ph} \rangle_\alpha \!-\! \frac{1}{2}\left[ \langle S_{\rm sp-ph}^2 \rangle_\alpha 
    \!-\! \langle S_{\rm sp-ph} \rangle^2_\alpha\right]
\end{eqnarray}
where $\langle ... \rangle_\alpha = \int D[\bar{\alpha}\alpha](...)e^{-S_{\rm sp}}$ with the expectation value taken using the spin-only 
action. Here, the bare phonon action $S^{\rm ph}_0$ is given by
\begin{eqnarray}
    S^{\rm ph}_0 &=& \sum_\br\! \sum_\ell \int_0^\beta\!\!\!\!\! d\tau~ {b}^*_{\br \ell}(\tau)(\partial_\tau + \Omega_0) b^\pdg_{\br \ell}(\tau)
\end{eqnarray}
and the spin-phonon interaction is given by
\begin{eqnarray}
    &S&_{\rm sp-ph} \!\!=\!\! \lambda \! \sum_\br\!\! \int_0^\beta \!\!\!\! d\tau [(b^\pdg_{\br x}(\tau)\!+\! b^*_{\br x}(\tau))(\sqrt{2 S}\sin\phi_\br(\alpha^\pdg_{\br,\tau} \!+\! \alpha^*_{\br,\tau})  \nonumber \\
    &+& \cos\phi_\br(2 \sin\theta_\br(S - \alpha^*_{\br,\tau} \alpha^\pdg_{\br,\tau})-i\sqrt{2 S}\cos\theta_\br(\alpha^\pdg_{\br,\tau} - \alpha^*_{\br,\tau})))  \nonumber \\
    &+&(b^\pdg_{\br z}(\tau)+b^*_{\br z}(\tau))(2 \cos\theta_\br(S - \alpha^*_{\br,\tau} \alpha^\pdg_{\br,\tau})
     \nonumber \\ 
     &+& i\sqrt{2 S}\sin\theta_\br(\alpha^\pdg_{\br,\tau} - \alpha^*_{\br,\tau}))].
\end{eqnarray}
where $\tau$ in these expressions refers to imaginary time. The first term in perturbation theory yields
a term linear in the phonon displacement which we absorb into a static coordinate shift.
Going to second order terms in the cumulant expansion (see Supplementary Material (SM) \cite{suppmat} for details), the effective phonon action 
is
\begin{equation}
    S^{\rm ph}_{\rm eff} = \sum_{\br,n} \begin{pmatrix}b_{\br x}^*(n) & b_{\br z}^*(n)\end{pmatrix}
        \cdot {\cal G}^{-1}(\br,i\nu_n) \cdot
    \begin{pmatrix} b_{\br x}(n)\\ b_{\br z}(n)\end{pmatrix}
    \label{eq:Sph}
\end{equation}
where $n$ labels the Matsubara frequency $i\nu_n = 2 n \pi T$, and 
${\cal G}^{-1}(\br,i\nu_n)$ is the $2\times 2$ inverse local phonon Green function matrix which depends on the 
site $\br$ in a general disordered spin background.
Explicitly,
\begin{eqnarray}
\label{eq:Ginv1}
    \!\!\!\!\! {\cal G}^{-1}_{xx}(\br,i\nu_n) \!&=&\! -i\nu_n\!+\!\Omega_0 -4S\lambda^2\left[ \frac{2h_\br\Delta^x_\br}{(i \nu_n)^2- 4h_\br^2} \right ]
    \\
    \label{eq:Ginv2}
    \!\!\!\!\! {\cal G}^{-1}_{zx}(\br,i\nu_n) \!&=&\! {\cal G}^{-1}_{xz}(\br,-i\nu_n) \!=\!  -4S\lambda^2 \! \left[\frac{2h_\br \chi_\br \!+\!i\nu_n \gamma_\br}{(i \nu_n)^2-4 h_\br^2} \right ]
    \\
    \label{eq:Ginv3}
    \!\!\!\!\! {\cal G}^{-1}_{zz}(\br,i\nu_n) \!&=&\! -i\nu_n+\Omega_0-4S\lambda^2\left[ \frac{2h_\br \Delta^z_\br}{(i\nu_n)^2- 4 h_\br^2} \right ]
\end{eqnarray}
where $\Delta^x_\br\!=\!\sin^2\phi_\br\!+\!\cos^2\phi_\br\cos^2\theta_\br
$, $\Delta^z_\br\!=\! \sin^2\theta_\br$,
$\chi_\br\!=\! -\sin2\theta_\br\cos\phi_\br$,
$\gamma_\br\!=\! 2i\sin\theta_\br\sin\phi_\br$ and $S=1/2$. 

This is the key result of our work, which
incorporates quantum fluctuations of the pseudospins as
well as their thermal fluctuations via Monte Carlo averaging of the pseudospin background.
As special cases, let us consider pure octupolar and pure quadrupolar orders. For octupolar order,
the pseudospin points along the $\tau_y$ axis, so $\theta\!=\!\pi/2$, $\phi\!=\!\pm \pi/2$, which
leads to $\Delta^x=\Delta^z=1$, $\gamma=\pm 2i$, and $\chi=0$.
Analytically continuing to the real frequency axis, the imaginary off-diagonal component in ${\cal G}^{-1}$
leads to eigenvectors which are {\it pseudo-chiral} superpositions $Q_x\pm iQ_z$ with nonzero PAM.
Conversely, when considering quadrupolar order, $\gamma=0$ and the
eigenmodes are {\it real} superpositions of $Q_x,Q_z$ which do not carry PAM. Turning to the spatial
structure of phonon eigenmodes, FO order would host uniform phonon pseudo-chirality while AFO order would
exhibit staggered phonon pseudo-chirality. Similarly, for FQ versus AFQ orders, we expect uniform versus
staggered real superpositions of $Q_x,Q_z$.


\noindent{\bf Phonon spectral function:}
Since the phonons are Einstein modes, we invert ${\cal G}^{-1}(\br,i\nu_n)$ at each site to obtain the local phonon Green function.
Focusing on the zero momentum phonon modes which can be probed in optics, e.g. via Raman spectroscopy \cite{Ye2019b}, 
we analytically continue $i\nu_n \to \omega+i 0^+$ to get
the phonon spectral function $A(\omega)=-(1/\pi) \sum_\br {\rm Tr} {\rm Im}[{\cal G} (\br,\omega)]$. 
Fig.\ref{fig:specFuncs-local} shows $A(\omega)$, averaged over $\sim\!10^5$
MC spin 
configurations, plotted as a function of frequency at various temperatures for no phonon damping
as well as phonon damping implemented by $\Omega_0 \! \to\! \Omega_0 \!+\! i \eta$ with a 
realistic value $2\pi\eta/\Omega_0\sim 0.07$ \cite{luo2023large,maehrlein2017terahertz}.
We have chosen Hamiltonian parameters to explore both the FO and AFQ phases.



Deep in the FO phase $T \! \ll \! T_{\rm FO}$, the pseudospin configurations are nearly uniform and polarized along $\tau_y$. 
Fig.\ref{fig:specFuncs-local}(a) shows the spectral function in the absence of phonon damping, with two split peaks 
in $A(\omega)$ associated with broken time-reversal symmetry in the FO state. These peaks at
$\Omega_{\pm}$, the advertised {\it pseudo-chiral} 
phonon modes, are split by
\textcolor{black}{$\Delta\Omega \!\!\approx\!\! 4 \lambda^2/\Omega_0 \!\!\approx\!\! 1.5$\,meV} due to the $\gamma_\br$ term in the off-diagonal 
component ${\cal G}_{zx}^{-1}$ in Eq.~\ref{eq:Ginv2}. 
However, they are not split symmetrically about $\Omega_0$; the diagonal $\Delta^x_\br,\Delta^z_\br$ terms 
in ${\cal G}^{-1}(\br,i\nu_n)$ lead to a downward shift (redshift) of the center frequency from \textcolor{black}{$\Omega_0$ by 
$4 \lambda^2 h_\br/\Omega_0^2 \approx 0.3$\,meV} where $h_\br\!\approx \!12 J_0$ for $T \ll T_{\rm FO}$.



With increasing temperature, spatial fluctuations of the pseudospins lead to a filling-in of the spectral gap in $A(\omega)$,
as the Einstein phonons at different sites in the system sense slightly different environments over the average octupolar order.
Sites within an octupolar domain will sense a large FO Weiss field, whereas sites in the
vicinity of domain walls where the Weiss field $\boldsymbol{h}(\br)$ vanishes will lead to
zero splitting. Averaging over these local environments causes a spread in the peak splittings, reflected in the 
average $A(\omega)$ in Fig.~\ref{fig:specFuncs-local}(a).
Finally, while sharp spectral peaks signalling
a long-range ordered FO phase are absent for $T \!>\! T_{\rm FO}$, there is nevertheless a `pseudogap' regime where
two split broad peaks at $\Omega_{\pm}$, with spectral weight $10\%$ more than that at $\omega\!=\!0$, persist to a
higher temperature $T^* \! \sim \! 1.2~T_{\rm FO}$ (see SM \cite{suppmat} for more details).

Fig.~\ref{fig:specFuncs-local}(b) shows the temperature dependent $A(\omega)$ incorporating realistic phonon 
damping $2\pi\eta/\Omega_0 \sim 0.07$ \cite{luo2023large,maehrlein2017terahertz}.
Notably, we see that split peaks in the FO phase remain resolvable despite the broadening, which could serve as a 
signature of FO order. However, the large 
phonon linewidth masks the `pseudogap' regime, so $A(\omega)$ exhibits a single peak at $\omega=\Omega_0$ for $T > T_{\rm FO}$.

We contrast these spectra in the FO phase with the corresponding spectra in the AFQ phase shown in Figs.~\ref{fig:specFuncs-local}(c) and (d).
In this case, the spectral splitting arises from ordering along $\tau_x,\tau_z$ and involves breaking of point group symmetry rather than
time-reversal symmetry.
In contrast to the FO case, the mode splitting in the AFQ phase is governed by unequal {\it diagonal} 
terms $\Delta^x$ and $\Delta^z$ in $\mathcal{G}^{-1}(\br,i\nu_n)$ in Eqs.~\ref{eq:Ginv1} and \ref{eq:Ginv3}. This leads to one mode at $\Omega_0$ and the second 
mode shifted downward (redshifted) to $\Omega_0-\Delta\Omega$ with
the mode splitting \textcolor{black}{$\Delta \Omega \!\approx\! 4\lambda^2h_\br/\Omega_0^2\!\approx\!0.4$\,meV} 
since $h_\br \!=\! (6\!-\!2\gamma)J_1 \!\approx\! 6.8 J_1$ deep in the AFQ phase. 
This mode splitting in the AFQ phase is parametrically smaller than that in the FO phase, being $\propto \! 1/\Omega_0^2$.
As a result, while a very weak splitting
is visible at low damping in Fig.~\ref{fig:specFuncs-local}(c), there is no visible peak splitting in the AFQ phase with realistic 
phonon damping in Fig.~\ref{fig:specFuncs-local}(d) even for $T \ll T_{\rm AFQ}$, but we observe a gradual $T$-dependent shift of the central 
peak.

The above results show that there can be important qualitative and quantitative differences in the phonon spectral function in the FO and 
AFQ phases. Probing the phonons may thus provide a route to discern between these different multipolar orders in candidate materials.

\begin{figure}[t]
\includegraphics[width=0.48\textwidth]{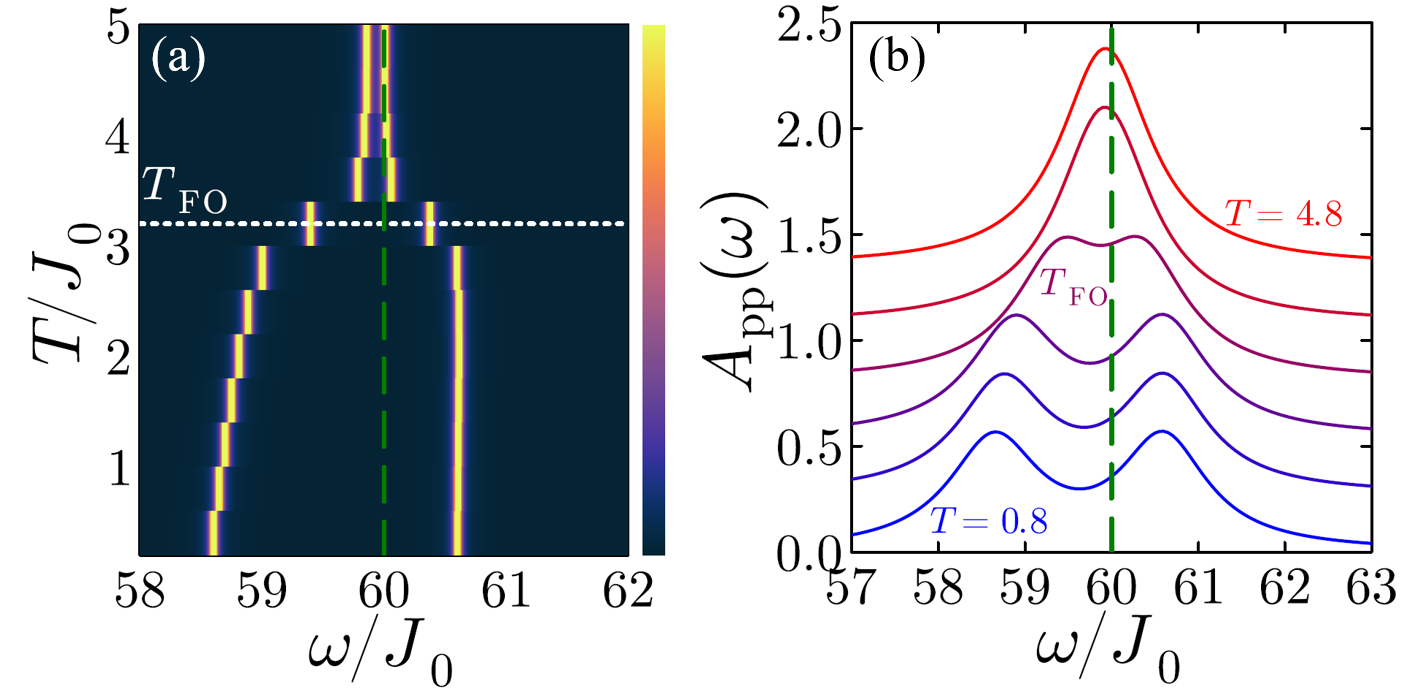}
\caption{\textcolor{black}{Phonon spectral function relevant to pump-probe experiments
$A_{\rm pp}(\omega)$ (see text) at equally spaced temperatures 
for $J_1=1.3J_0$ which leads to a FO phase as in Fig.~\ref{fig:specFuncs-local}.
For $T < T_{\rm FO}$, there are two distinct sharp peaks in the (a) undamped system, while a realistic damping 
parameter of  (b) $2\pi \eta/\omega = 0.07$ broadens the spectra but with resolvable two peaks
(successive $T$ curves are vertically offset by $0.27$ for clarity). The splitting vanishes above $T_{\rm FO}$ where it merges into the 
\textcolor{black}{natural phonon frequency $\Omega_0 = 60 J_0$} (green dotted line).} 
}
\label{fig:specFuncs-q=0}
\end{figure}

\noindent{\bf Pumped phonons:} The action derived in Eqns.~\ref{eq:Sph}-\ref{eq:Ginv3} is also useful to understand optical pump-probe
spectroscopy of phonon modes. A pump pulse can lead to an effective large coherent phonon occupation of only the $\bk=0$ mode, so
we can replace this phonon mode by an effective classical field. Ignoring all modes with
momenta $\bk \neq 0$ for short-time dynamics, the effective action is
\begin{equation}
    \!\! S^{\rm ph,0}_{\rm eff} \!=\!\!\!\! \sum_{n} \!\! \begin{pmatrix}b_{0 x}^*(n) \!&\! b_{0 z}^*(n)\end{pmatrix}
        \! \cdot \! \left[{\cal G}_{\rm pp}^{-1}\!(i\nu_n)\right] \!\cdot\!
    \begin{pmatrix} b_{0 x}(n)\\ b_{0 z}(n)\end{pmatrix}
    \label{eq:Sph0}
\end{equation}
where ${\cal G}_{\rm pp}^{-1}\!(i\nu_n) = \frac{1}{N} \sum_\br\! {\cal G}^{-1}\!(\br,i\nu_n)$. Fig.~\ref{fig:specFuncs-q=0} plots the
spectral function 
$A_{\rm pp}(\omega) = - (1/\pi) {\rm Tr Im} {\cal G}_{\rm pp}\!(\omega)$
incorporating realistic phonon damping as explained earlier. This result involves an effective spatial averaging
of ${\cal G}^{-1}\!(\br,i\nu_n)$ rather than ${\cal G}(\br,i\nu_n)$ as done earlier,
so $A_{\rm pp}(\omega)$ for zero
damping exhibits no pseudogap regime; rather the split peaks merge for $T > T_{\rm FO}$. 
The split peaks observed for $T < T_{\rm FO}$
correspond to modes $Q_x \pm i Q_z$ which can manifest as `beat dynamics' in the FO phase.
Exciting the $Q_x$ mode with a pump pulse can lead to energy transfer into the $Q_z$ mode over a
timescale $2\pi/\Delta\Omega$ as observed in simulations \cite{hart2024phonon}, while scattering
to $\bk \neq 0$ modes will cause additional damping at longer timescales. For $T > T_{\rm FO}$, the spectral gap vanishes,
leading to a single peak at $\Omega_0$.

\begin{figure}[t]
\includegraphics[width=0.48\textwidth]{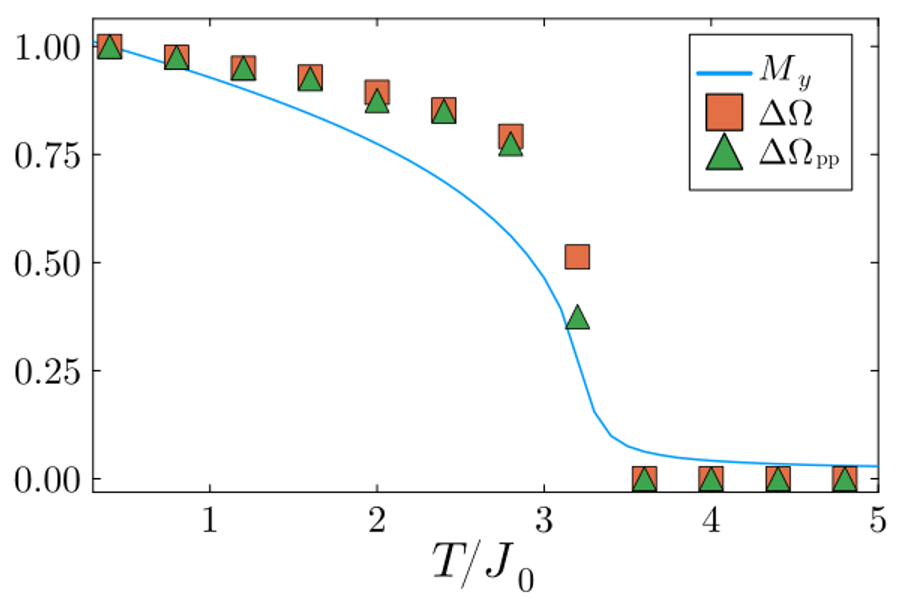}
\caption{\textcolor{black}{Temperature dependence of the pseudo-chiral phonon splitting with damping of $2\pi\eta/\Omega_0 \sim 0.07$, extracted from the phonon spectral function (orange squares) and the pump-probe spectral function (green triangles), normalized to their value at $T=0$. 
For comparison, we show the octupolar order parameter $M_y(T) = \frac{1}{N}
|\sum_{\br}\langle \tau_{\br y} \rangle|$ (solid blue line).}}
\label{fig:splitting-OP}
\end{figure}

\noindent {\bf Discussion:}
In this work, we propose a new probe for multipolar magnetic order which leverages symmetry-breaking imprint on phonons. While recent works on dipolar magnets~\cite{wu2025magnetic, Che2025PRL} have observed the signatures of phonon chirality and mode splitting via Raman spectroscopy, 
the phonon signatures of multipolar magnetism remain unexplored. 
Using  a path-integral approach, we show a FO order induced pseudochiral phonons with a splitting $\sim\!1$\,meV, similar to values reported
for dipolar magnets \cite{wu2025magnetic, Che2025PRL}.
These pseudo-chiral phonons
can be observed via Raman spectroscopy in the back scattering geometry with light propagating along the cubic [111] axes, with different
intensities in different circularly polarized channels (see \cite{suppmat} for details).
To the best of our knowledge, this is the first such prediction,  which does not rely on magnetic field, net magnetization, 
non-trivial band topology, or even on the circular motion of ions!

Interestingly, while our results for the normalized T-dependent mode splitting 
qualitatively tracks, but is not simply proportional to, the FO order
as highlighted in Fig.\ref{fig:splitting-OP}.
We trace this to the fact that the splitting in the FO phase is governed by the $\gamma_\br$ off-diagonal term in ${\cal G}^{-1}(\br,i\nu_n)$, 
which tracks only
the \emph{direction} but not the magnitude of the local Weiss fields. A similar discrepancy between mode splitting and the
magnetic order parameter
in seen in recent experimental results on the ferrimagnetic insulator {Fe$_{1.75}$Zn$_{0.25}$Mo$_3$O$_8$} \cite{wu2025magnetic}. We believe that our 
approach can capture such effects as arising from spatial non-uniformities of the order parameter. 

In summary, our phonon-based proposal offers an accessible route to probing octupolar orders with optical spectroscopy, 
complementing other tools such as magnetostriction \cite{patri2019unveiling,patri2020theory}, polarized neutron scattering \cite{Lovesey2020}, 
NMR \cite{tokunaga2006nmr,LiuW2018,Cong2023,voleti2023probing}, and X-ray diffraction
\cite{hirai2020detection}. Using such complementary probes could shed valuable light on complex multipolar orders in 
a wide variety of quantum materials. Moreover, the theoretical approach used here can accommodate more general TRS breaking scenarios and might find applications in a wide variety of antiferromagnets, 
altermagnets \cite{Smejkal2022,jungwirth2024,jungwirth2025_ALMreview}, 
or even in the pseudo-gap phase of cuprates where thermal Hall measurements have hinted at chiral phonons~\cite{grissonnanche2020chiral,Grissonnanche2019}.

\noindent {\it Note added:} Recently, we became aware of a related work~\cite{watanabe2025dual} that proposes dual-circular Raman optical activity as a probe for axial multipolar order.

\noindent {\bf Acknowledgments:}
We thank Bruce Gaulin for inspiring discussions.
This research was supported by Discovery Grant RGPIN-2021-03214 from the Natural Sciences and Engineering Council (NSERC) of Canada (A.P.),
an NSERC CGS-D fellowship (R.S.), and an Ontario Graduate Scholarship (K.H.).
S.C. acknowledges support from  JSPS KAKENHI No. JP23H04865, MEXT, Japan. This research was supported in part by the International Centre for Theoretical Sciences (ICTS) for the program - Engineered 2D Quantum Materials (code: ICTS/E2QM2024/07). Numerical computations were performed on the Niagara supercomputer at the SciNet HPC 
Consortium and the Digital Research Alliance of Canada.

\bibliography{main.bib, ref2.bib, heavyfermions.bib}

\end{document}


\preprint{APS/123-QED}
\title{Supplementary Material: \\ Pseudo-chiral phonon splitting from octupolar magnetic order}
\author{Ruairidh Sutcliffe}
\email{ruairidh.sutcliffe@mail.utoronto.ca}
\affiliation{Department of Physics, University of Toronto, 60 St. George Street, Toronto, ON, M5S 1A7 Canada}
\author{Kathleen Hart}
\email{kf.hart@mail.utoronto.ca}
\affiliation{Department of Physics, University of Toronto, 60 St. George Street, Toronto, ON, M5S 1A7 Canada}
\author{Swati Chaudhary}
\email{swatichaudhary@issp.u-tokyo.ac.jp}
\affiliation{The Institute for Solid State Physics, The University of Tokyo, Kashiwa, Chiba 277-8581, Japan}
\author{Arun Paramekanti}
\email{arun.paramekanti@utoronto.ca}
\affiliation{Department of Physics, University of Toronto, 60 St. George Street, Toronto, ON, M5S 1A7 Canada}

\date{\today}

\maketitle

\beginsupplement

\section{$E_g$ phonons in $\rm Ba_2MgOsO_6$}

For the purpose of calculating displacement vectors associated with $E_g$ phonons, we consider double perovskite Ba\textsubscript{2}MgOsO\textsubscript{6}. It crystallizes in a face-centered cubic (FCC) structure with space group \textbf{Fm$\bar{3}$m} (No. 225). This structure is characterized by the rock-salt ordering of the B-site cations (Mg\textsuperscript{2+} and Os\textsuperscript{6+}), embedded within a framework of corner-sharing oxygen octahedra. The parent perovskite structure has a general formula of A\textsubscript{2}BB'O\textsubscript{6}, where:
\begin{itemize}
  \item The A-site is occupied by Ba\textsuperscript{2+} cations.
  \item The B and B' sites are occupied by Mg\textsuperscript{2+} and Os\textsuperscript{6+} respectively, in a rock-salt type arrangement.
  \item Oxygen atoms form a three-dimensional octahedral network around each B-site cation.
\end{itemize}

The atomic positions in the unit cell are described by the following Wyckoff positions:

\begin{itemize}
  \item Ba\textsuperscript{2+}: occupies the $8c$ site at coordinates $(\tfrac{1}{4}, \tfrac{1}{4}, \tfrac{1}{4})$
  \item Mg\textsuperscript{2+}: occupies the $4a$ site at coordinates $(0, 0, 0)$
  \item Os\textsuperscript{6+}: occupies the $4b$ site at coordinates $(\tfrac{1}{2}, \tfrac{1}{2}, \tfrac{1}{2})$
  \item O\textsuperscript{2-}: occupies the $24e$ site at coordinates $(x, 0, 0)$, where $x \approx 0.24$. These oxygen ions are shown by six different colors in Fig.~\ref{fig:Egmode}.
\end{itemize}
\subsection*{Displacement pattern for $E_g$ modes}

The $E_g$ mode involves distortions of the oxygen octahedra, particularly planar stretching and compressing motions. We consider a basis consisting of six oxygen ions located at the positions $(x, 0, 0)$, $(-x, 0, 0)$, $(0, x, 0)$, $(0, -x, 0)$, $(0, 0, x)$, and $(0, 0, -x)$, which undergo displacements in different directions. The remaining eighteen oxygen ions in the unit cell can be generated by translating these six basis atoms by the vectors $(0, \tfrac{1}{2}, \tfrac{1}{2})$, $(\tfrac{1}{2}, 0, \tfrac{1}{2})$, and $(\tfrac{1}{2}, \tfrac{1}{2}, 0)$. The corresponding displacements of these translated atoms follow from those of the original six.

\begin{figure}
\includegraphics[width=1.0\linewidth]{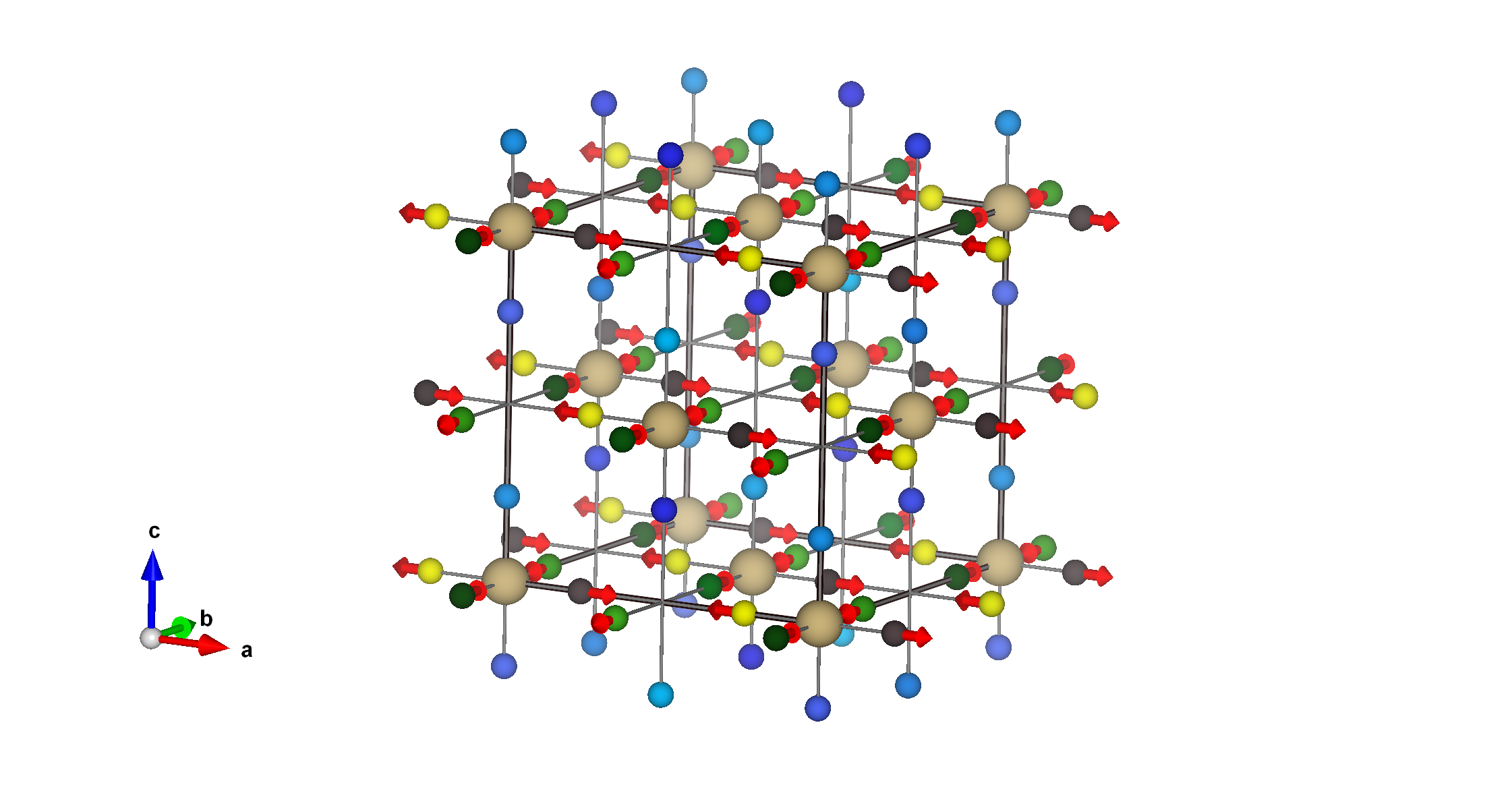}
\includegraphics[width=1.0\linewidth]{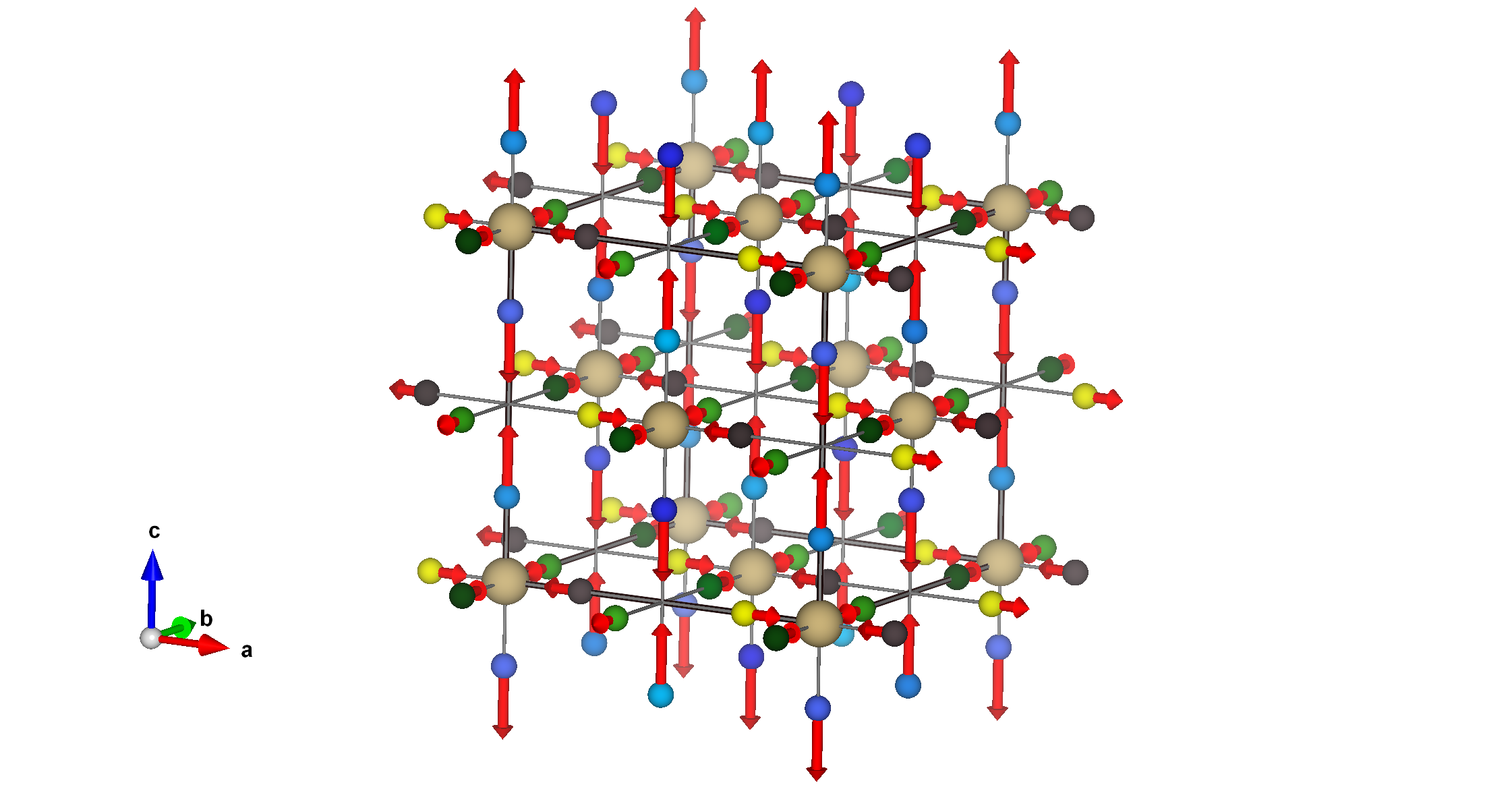}
\caption{Displacement patterns for the two components of the ${\cal E}_g$ mode. The six oxygen ions are shown in different colors. For clarity, Mg and Ba atoms are omitted.
}
\label{fig:Egmode}
\end{figure}

To clearly illustrate the displacement pattern, the six oxygen atoms in each Os-centered octahedron are shown in six different colors in Fig.~\ref{fig:Egmode}. Using the transformation properties of the irreducible representation $E_g$, we obtain the following displacement pattern:

\begin{equation}
    Q(E_g^a)=\frac{1}{2}\left(\hat{x},-\hat{x},-\hat{y},\hat{y},0,0\right)
\end{equation}
\begin{equation}
    Q(E_g^b)=\frac{1}{2\sqrt{3}}\left(-\hat{x},\hat{x},-\hat{y},\hat{y},2\hat{z},-2\hat{z}\right)
\end{equation}
in the basis of six oxygen ions (O$^1$,O$^2$,O$^3$,O$^4$,O$^5$,O$^6$) located at the positions $(x, 0, 0)$, $(-x, 0, 0)$, $(0, x, 0)$, $(0, -x, 0)$, $(0, 0, x)$, and $(0, 0, -x)$, respectively. These displacement patterns are shown in Fig.~\ref{fig:Egmode}. The most interesting feature to note here is that both components exhibit linear motion in the same direction. This implies that the chiral superposition $Q(E_g^a) \pm i Q(E_g^b)$ cannot result in circular motion. However, it has been recently shown in Ref.~\cite{chaudhary2025anomalous} that even such rotationless phonons can carry quantized pseudo-angular momentum (PAM). It is defined in terms of the phase picked up by the phonon eigenstate under three-fold rotation and has been commonly employed to study chiral valley phonons~\cite{Zhang2015PRL}. In a crystal with $m$-fold rotational symmetry about $\hat{n}$ axis, for a phonon mode $\nu$ with momentum $\mathbf{q}$, PAM $\mathbf{l}^p_{\nu\mathbf{q}}$ points along $\hat{n}$ is defined as
\begin{equation}
   C_m^{\hat{n}}\mathbf{u}_{\nu\mathbf{q}}e^{i\mathbf{R}_{\alpha l} \cdot\mathbf{q}} =e^{-i\frac{2\pi}{m}\hat{n}\cdot\mathbf{l}^p_{\nu\mathbf{q}}}\mathbf{u}_{\nu\mathbf{q}} e^{i\mathbf{R}_{\alpha l} \cdot\mathbf{q}},
    \label{Eq:rotationOAM}
\end{equation}
and can take values of $l^p_{\nu\mathbf{q}}=0,...,(m-1)$, and $\mathbf{R}_{\alpha l}$ is the position vector for atom $\alpha$ in unit cell $l$. The phase factor can arise from a rotation of the displacement vector $\mathbf{u}_{\nu\mathbf{q}}$ directly (intracell, spin PAM), or from a rotation of the nonlocal part, $e^{i\mathbf{R}_j \cdot\mathbf{q}}$, (intercell, orbital PAM). This quantized PAM arises at points in the Brillouin zone that respect $n$-fold rotational symmetry.

In the current scenario of ${\cal E}_g$ modes in a double-perovskite crystal, the point group of the crystal has three-fold rotational symmetry about each diagonal of the cubic unit cell. We can define PAM according to the transformation of chiral ${\cal E}_g$ modes under a three-fold rotation about the following four-diagonals 
\begin{equation}
    \hat{n}_1=\frac{1}{\sqrt{3}}(\hat{x}+\hat{y}+\hat{z}),\, \hat{n}_2=\frac{1}{\sqrt{3}}(\hat{x}+\hat{y}-\hat{z})
    , \hat{n}_3=\frac{1}{\sqrt{3}}(\hat{x}+\hat{y}-\hat{z}),\, \hat{n}_4=\frac{1}{\sqrt{3}}(\hat{x}-\hat{y}-\hat{z}).
\end{equation}
The chiral eigenmodes are given by:
\begin{equation}
  Q^{\pm}=\frac{1}{\sqrt{2}}\left(Q(E_g^a)\pm iQ(E_g^b)\right)=\frac{i}{\sqrt{3}}\left(\hat{x}\,e^{\pm i\frac{4\pi}{3}},-\hat{x}\,e^{\pm i\frac{4\pi}{3}},\hat{y}\,e^{\pm i\frac{2\pi}{3}},-\hat{y}\,e^{\pm i\frac{2\pi}{3}},\hat{z},-\hat{z}\right)
\end{equation}
and they satisfy: 
\begin{equation}
    C_3^{\hat{n}_1}Q^{\pm}=e^{\pm i \frac{2\pi}{3}}Q^{\pm},\, C_3^{\hat{n}_2}Q^{\pm}=e^{\mp i \frac{2\pi}{3}}Q^{\pm},\,
    C_3^{\hat{n}_3}Q^{\pm}=e^{\mp i \frac{2\pi}{3}}Q^{\pm},\,
    C_3^{\hat{n}_4}Q^{\pm}=e^{\pm i \frac{2\pi}{3}}Q^{\pm}
\end{equation}
which indicates that $Q^{\pm}$ carry opposite angular momentum when measured about one of the diagonal $\hat{n}_i$ of the cubic unit cell. Additionally, the sign of this pseudo-angular momentum also depends on the direction of rotation axis. This kind of PAM is important for angular momentum conservation in Raman scattering and selection rules~\cite{Zhang2015PRL,Zhu2018,Chen2019_entangle,ueda2023chiral,ishito2023truly}.

\subsection*{Raman tensor and selection rules}
In the $(x,y,z)$ crystal frame, Raman tensor for ${\cal E}_g$ modes  is given by:
\begin{equation}
    R{({\cal E}_g^a)} =
\begin{pmatrix}
\sqrt{3}\,a & 0 & 0 \\
0 & -\sqrt{3} \,a & 0 \\
0 & 0 & 0
\end{pmatrix}, 
\qquad
R{({\cal E}_g^b)} =
\begin{pmatrix}
-a & 0 & 0 \\
0 & -a & 0 \\
0 & 0 & 2a
\end{pmatrix},
\end{equation}
where $a \equiv a(\omega)$ is a function of frequency. 
As discussed above, a chiral superposition of these modes carry PAM along the diagonal of the cube. In order to detect the angular momentum in Raman scattering, we can consider light propagating along $\hat{n}_1$, i.e $[111]$ direction. Next, we can define a set of three cartesian vectors with their axis aligned along  $\hat{n}_1$: 
\begin{equation}
    \hat{e}_1 = \tfrac{1}{\sqrt{2}}(1,-1,0), \quad
\hat{e}_2 = \tfrac{1}{\sqrt{6}}(1,1,-2), \quad
\hat{e}_3 = \tfrac{1}{\sqrt{3}}(1,1,1).
\end{equation}

In this basis Raman tensor for two components of the given ${\cal E}_g$ mode takes the following form:
\begin{equation}
    R'{({\cal E}_g^a)} =
\begin{pmatrix}
0 & a & \sqrt{2}\,a \\
a & 0 & 0 \\
\sqrt{2}\,a & 0 & 0
\end{pmatrix}, 
\qquad
R'{({\cal E}_g^b)} =
\begin{pmatrix}
-a & 0 & 0 \\
0 & a & -\sqrt{2}\,a \\
0 & -\sqrt{2}\,a & 0
\end{pmatrix}.
\end{equation}
resulting in the following Raman tensors for $Q^\pm$ modes:
\begin{equation}
    R'{(Q^+)} =
\begin{pmatrix}
-i\,a & a & \sqrt{2}\,a \\
a & i\,a & -i\,\sqrt{2}\,a\, \\
\sqrt{2}\,a & -i\,\sqrt{2}\,a\ & 0
\end{pmatrix}, 
\qquad
R'{(Q^-)} =
\begin{pmatrix}
i\,a & 0 & \sqrt{2}\,a \\
0 & -i\,a & i\,\sqrt{2}\,a \\
\sqrt{2}\,a & i\,\sqrt{2}\,a & 0
\end{pmatrix}.
\end{equation}

When time-reversal symmetry is broken, the frequency dependence of two Raman channels can be different, and we can write:
\begin{equation}
    R'{(Q^+)} =
\begin{pmatrix}
-i\,a^+ & a^+ & \sqrt{2}\,a^+ \\
a^+ & i\,a^+ & -i\,\sqrt{2}\,a^+\, \\
\sqrt{2}\,a^+ & -i\,\sqrt{2}\,a^+\ & 0
\end{pmatrix}, 
\qquad
R'{(Q^-)} =
\begin{pmatrix}
i\,a^- & 0 & \sqrt{2}\,a^- \\
0 & -i\,a^- & i\,\sqrt{2}\,a^-\\
\sqrt{2}\,a^- & i\,\sqrt{2}\,a^- & 0
\end{pmatrix}.
\end{equation}
where $a^\pm \equiv a^\pm(\omega)$ will be peaked at respective frequencies $\Omega_\pm$.

In the back scattering geometry, and for light propagating along $\hat{e}_3$, intensities in different circularly polarized channels satisfy the following relationship:
\begin{equation}
    |\sigma^\pm\cdot  R'{(Q^+)}\cdot \sigma^\pm|^2=0;  \quad|\sigma^+\cdot  R'{(Q^+)}\cdot \sigma^-|^2=|a^+|^2 
\end{equation}
\begin{equation}
    |\sigma^\pm\cdot  R'{(Q^-)}\cdot \sigma^\pm|^2=0;\quad  |\sigma^-\cdot  R'{(Q^-)}\cdot \sigma^+|^2=|a^-|^2 
\end{equation}
where $\sigma^\pm$ refers to left and right circular polarization. This suggests that phonons with opposite PAM appear in opposite cross-circular channels
and will be peaked at corresponding frequencies $\Omega_\pm$ which are split below the octupolar ordering transition.

\section{Details of the derivation of the effective phonon action}
In this section, we provide a discussion of the effective phonon action, as discussed in the main text. Explicitly, we may begin with the spin-phonon Hamiltonian described in the main text, namely:
\begin{eqnarray}
    \!\!\!\!\!\!\!\!  H_{\rm sp-ph}= \! \lambda \sum_\br \Big\{&(b_{\br x}\!+\! b^\dagger_{\br x})&\big[\sqrt{2 S}\sin\phi_\br(\alpha_\br \!+\! \alpha^\dagger_\br)
    \!\!+ \! 2 \sin\theta_\br\cos\phi_\br(S \!-\! \alpha^\dagger_\br \alpha_\br)
    \!+\! i\sqrt{2 S}\cos\theta_\br\cos\phi_\br(\alpha^\pdg_\br \!-\! \alpha^\dg_\br)\big] \nonumber \\
    \!\!+\! &(b^\pdg_{\br z}\!\!+\! b^\dg_{\br z})&\big[2 \cos\theta_\br(S \!-\! \alpha^\dagger_\br \alpha_\br)
    \!+\! i\sqrt{2 S}\sin\theta_\br(\alpha^\pdg_\br \!-\! \alpha^\dg_\br)\big]\Big\}
\end{eqnarray}
where $\phi_\br,\theta_\br,h_\br$ describe the local field direction of individual spins, determined from large-scale Monte Carlo simulations discussed in Ref.\cite{hart2024phonon}, and represent the azimuthal angle, polar angle and field amplitude, respectively. Thus, the full spin-phonon action may be written as 
\begin{eqnarray}
    S_{\rm sp-ph} \!\!=\!\! \lambda \! \sum_\br\!\! \int_0^\beta \!\!\!\! d\tau \Big\{&[&b^\pdg_{\br x}(\tau)\!+\! b^*_{\br x}(\tau)]\big[\sqrt{2 S}\sin\phi_\br(\alpha^\pdg_{\br,\tau} \!+\! \alpha^*_{\br,\tau})
    \!+\! 2 \sin\theta_\br\cos\phi_\br(S \!-\! \alpha^*_{\br,\tau} \alpha^\pdg_{\br,\tau}) \nonumber 
    \!-\! i\sqrt{2 S}\cos\theta_\br\cos\phi_\br(\alpha^\pdg_{\br,\tau}\!-\! \alpha^*_{\br,\tau})\big] 
    \\
    \!+\!&[&b^\pdg_{\br z}(\tau)\!+\!b^*_{\br z}(\tau)]\big[2 \cos\theta_\br(S \!-\! \alpha^*_{\br,\tau} \alpha^\pdg_{\br,\tau})  
     \!+\! i\sqrt{2 S}\sin\theta_\br(\alpha^\pdg_{\br,\tau} \!- \!\alpha^*_{\br,\tau})\big]\Big\}.
\end{eqnarray}
where, $\tau$ denotes imaginary time. We wish to compute the effective phonon action, integrating out the pseudospin degrees of freedom using a second-order cumulant expansion
\begin{equation}
    S^{ph}_{\rm eff} = S^{ph}_0 + \langle S_{sp-ph} \rangle_\alpha - \frac{1}{2}\left[ \langle S_{sp-ph}^2 \rangle_\alpha - \langle S_{sp-ph} \rangle^2_\alpha\right]
\end{equation}
where
\begin{equation}
    \langle S_{\rm sp-ph} \rangle_\alpha = \int D[\bar{\alpha}\alpha](S_{\rm sp-ph})e^{-S_{\rm sp}}
\end{equation}
where $S_{\rm sp}=\sum_\br \int_0^\beta d\tau \left[ {\alpha}^*_{\br,\tau}\left(\partial_\tau-E\right){\alpha}_{\br,\tau}\right]$ is the spin action and the spin energy is $E = 2h_\br$, as defined in the main text.

The linear term in the expansion is linear in phonon displacement which can be absorbed into the phonon action $S_0^{\rm ph}$ as a displacement shift, such that $\tilde{S}_0^{\rm ph} = S^{\rm ph}_0 + \langle S_{\rm sp-ph} \rangle_\alpha$; see main text for details. Next, we compute the second order terms $S_2 = - \frac{1}{2}\left[ \langle S_{\rm sp-ph}^2 \rangle_\alpha - \langle S_{\rm sp-ph} \rangle^2_\alpha\right]$, which we break up into four terms $S_2 = S_{2}^{xx} + S_{2}^{xz} + S_{2}^{zx} + S_{2}^{zz}$. After some simplification and writing in terms of the spin Green functions
\begin{equation}
 \mathcal{G}_\br^S(\tau-\tau')=\left<\mathcal{T}\alpha_{\br,\tau}\alpha^*_{\br,\tau'}\right>, 
\end{equation}
our second order expansion of the effective phonon action is of the form (Note that the expectation value of alphas fixes $\br=\br'$),
\begin{eqnarray}
    S_{2}^{xx} \!=\!  -\!\lambda^2\sum_\br\int_0^\beta d\tau d\tau'  \left[b_{\br x}(\tau)\!+\!b^*_{\br x}(\tau)\right] \!&\Bigl\{&\! \left[{S}\sin^2\phi_\br \!+\! {S}\cos^2\theta_\br\cos^2\phi_\br\right]\left[\mathcal{G}_\br^S(\tau\!-\!\tau') \!+\! \mathcal{G}_\br^S(\tau'\!-\!\tau)\right] \nonumber
    \\\!&+&\! 2 \sin^2\theta_\br\cos^2\phi_\br\mathcal{G}_\br^S(\tau'\!-\!\tau)\mathcal{G}_\br^S(\tau\!-\!\tau')\Bigl\}\left[b_{\br x}(\tau')\!+\!b^*_{\br x}(\tau')\right]
\\
    S_{2}^{xz} \!=\!  -\!\lambda^2\sum_\br\int_0^\beta d\tau d\tau' \left[b_{\br x}(\tau)\!+\!b^*_{\br x}(\tau)\right] \!&\Bigl\{&\! i{S}\sin\phi_\br\sin\theta_{\br}\left[\mathcal{G}^S_\br(\tau'\!-\!\tau) \!-\! \mathcal{G}^S_\br(\tau\!-\!\tau')\right] \nonumber
    \\\!&+&\! 2 \sin\theta_\br\cos\phi_\br \cos\theta_{\br}\mathcal{G}^S_\br(\tau'\!-\!\tau) \mathcal{G}^S_\br(\tau\!-\!\tau')\nonumber
    \\\!&+&\! S\cos\theta_\br\cos\phi_\br\sin\theta_{\br}\left[\!-\!\mathcal{G}^S_\br(\tau\!-\!\tau') \!- \!\mathcal{G}^S_\br(\tau'\!-\!\tau)\right]\Bigl\} \left[b_{\br z}(\tau')\!+\!b^*_{\br z}(\tau')\right]
\\
    S_{2}^{zx} \!=\!  -\!\lambda^2\sum_\br\int_0^\beta d\tau d\tau' [b_{\br z}(\tau)\!+\!b^*_{\br z}(\tau)]\!&\Bigl\{&\!i{S}\sin\phi_{\br}\sin\theta_\br\left[\mathcal{G}^S_\br(\tau\!-\!\tau') \!-\! \mathcal{G}^S_\br(\tau'\!-\!\tau)\right] \nonumber\\
    \!&+&\! 2 \sin\theta_{\br}\cos\phi_{\br} \cos\theta_\br\mathcal{G}^S_\br(\tau'\!-\!\tau)\mathcal{G}^S_\br(\tau\!-\!\tau')\nonumber
    \\\!&+&\! {S}\cos\theta_{\br}\cos\phi_{\br}\sin\theta_\br \left[\!-\!\mathcal{G}^S_\br(\tau\!-\!\tau') \!-\! \mathcal{G}^S_\br(\tau'\!-\!\tau)\right]\Bigl\}[b_{\br x}(\tau')\!+\!b^*_{\br x}(\tau')]
\\
    S_{2}^{zz} \!=\!  -\!\lambda^2\sum_\br\int_0^\beta d\tau d\tau' [b_{\br z}(\tau)\!+\!b^*_{\br z}(\tau)]\!&\Bigl\{&\!2 \cos^2\theta_\br \mathcal{G}^S_\br(\tau'\!-\!\tau) \mathcal{G}^S_\br(\tau\!-\!\tau')\nonumber
    \\
    \!&-&\! {S}\sin^2\theta_\br\left[\!-\!\mathcal{G}^S_\br(\tau\!-\!\tau') \!-\! \mathcal{G}^S_\br(\tau'\!-\!\tau)\right]\Bigl\}[b_{\br z}(\tau')\!+\!b^*_{\br z}(\tau')]
\end{eqnarray}
Next, we use the Matsubara frequency representation of these Magnon Greens functions, i.e
\begin{equation}
    \mathcal{G}_\br^S(\tau-\tau')=\frac{1}{\beta}\sum_n \mathcal{G}^S_\br(i\nu_n)e^{-i\nu_n(\tau-\tau')};\,\, \nu_n=\frac{2n\pi}{\beta};\,\,\mathcal{G}^S_\br(i\nu_n)=\frac{1}{i\nu_n-2h_\br}
\end{equation}and also write down the phonon creation and annihilation operators in Matusbara space
\begin{equation}
    a_{\br \ell}(\tau)=\frac{1}{\sqrt{\beta}}\sum_n a_{\br \ell}(n)e^{-i\nu_n\tau},\,\, a^*_{\br \ell}(\tau)=\frac{1}{\sqrt{\beta}}\sum_n a^*_{\br \ell}(n)e^{i\nu_n\tau}
\end{equation}
Note that terms of the form $b_{\br \ell}(\tau)\mathcal{G}_\br^s(\tau'-\tau)\mathcal{G}^s_\br(\tau-\tau')b^*_{\br \ell}(\tau')$ will vanish since:
\begin{equation}
\int_0^\beta b_{\br \ell} (\tau) \mathcal{G}_\br^s(\tau'-\tau)\mathcal{G}^s_\br(\tau-\tau')b^*_{\br\ell}(\tau') =\sum_{n,n'}a_{\br\ell}(n)a^*_{\br\ell}(n)\mathcal{G}_\br^s(i\nu_{n'})\mathcal{G}_\br^s(i\nu_n+i\nu_{n'}) = \sum_n\frac{n_B(2h_\br - i\nu_n) - n_B(2h_\br)}{ i\nu_n},
\end{equation}
noting that $n_B(2h_\br - i\nu_n) = n_B(2h_\br)$. This leaves us with a rather simple form of the second order terms,
\begin{eqnarray}
    S_{2}^{xx} &=&  -\lambda^2 S\sum_{\br,n}\Delta^x_\br\left[\mathcal{G}^s_\br(i\nu_n)+\mathcal{G}^s_\br(-i\nu_n)  \right]\left[b_{\br x}(n) b^*_{\br x}(n) + b_{\br x}^*(n) b_{\br x}(n)\right]
\\
    S_{2}^{xz} &=&  -\lambda^2 S\sum_{\br,n}\left\{ \left[\kappa_\br\mathcal{G}^s_\br(i\nu_n)-\delta_\br\mathcal{G}^s_\br(-i\nu_n) \right]b_{\br x}(n)b^*_{\br z}(n)+\left[\kappa_\br\mathcal{G}^s_\br(-i\nu_n)-\delta_\br\mathcal{G}^s_\br(i\nu_n) \right]b^*_{\br x}(n)b_{\br z}(n)\right\}
\\
    S_{2}^{zx} &=&  -\lambda^2 S\sum_{\br,n}\left\{ \left[\kappa_\br\mathcal{G}^s_\br(i\nu_n)-\delta_\br\mathcal{G}^s_\br(-i\nu_n) \right]b_{\br x}(n)b^*_{\br z}(n)+\left[\kappa_\br\mathcal{G}^s_\br(-i\nu_n)-\delta_\br\mathcal{G}^s_\br(i\nu_n) \right]b^*_{\br x}(n)b_{\br z}(n)\right\}
\\
    S_{2}^{zz} &=&  -\lambda^2 S\sum_{\br,n}\Delta^z_\br\left[\mathcal{G}^s_\br(i\nu_n)+\mathcal{G}^s_\br(-i\nu_n) \right]\left[b_{\br z}(n) b^*_{\br z}(n) + b^*_{\br z}(n) b_{\br z}(n)\right]
\end{eqnarray}
where $\Delta^x_\br=\sin^2\phi_\br + \cos^2\theta_\br\cos^2\phi_\br$, $\Delta^z_\br=\sin^2\theta_\br$, $\kappa_\br = -\cos\theta_\br\cos\phi_\br\sin\theta_\br + i\sin\phi_\br\sin\theta_\br$ and $\delta_\br = \cos\theta_\br\cos\phi_\br\sin\theta_\br + i\sin\phi_\br\sin\theta_\br$.
Plugging in for $\mathcal{G}^s_\br$, 
\begin{eqnarray}
    S^{xx}_{2}&=&-\lambda^2\sum_{\br,n} \left[ \frac{2S\Delta^x_\br(2h_\br)}{(i\nu_n)^2-(2h_\br)^2} \right][b_{\br x}(n) b^*_{\br x}(n) + b_{\br x}^*(n) b_{\br x}(n)]
\\
    S^{xz}_{2}&=&-\lambda^2\sum_{\br,n} \left\{\left[\frac{2S\chi_\br (2h_\br)+S\gamma_\br(i\nu_n)}{(i\nu_n)^2-(2h_\br)^2} \right ]b_{\br x}(n)b^*_{\br z}(n)+\left[ \frac{2S\chi_\br (2h_\br)-S\gamma_\br(i\nu_n)}{(i\nu_n)^2-(2h_\br)^2}\right ]b^*_{\br x}(n)b_{\br z}(n)\right\}
\\
    S^{zx}_{2}&=&-\lambda^2\sum_{\br,n} \left\{\left[\frac{2S\chi_\br (2h_\br)+S\gamma_\br(i\nu_n)}{(i\nu_n)^2-(2h_\br)^2} \right ]b_{\br x}(n)b^*_{\br z}(n)+\left[ \frac{2S\chi_\br (2h_\br)-S\gamma_\br(i\nu_n)}{(i\nu_n)^2-(2h_\br)^2}\right ]b^*_{\br x}(n)b_{\br z}(n)\right\}
\\
    S^{zz}_{2}&=&-\lambda^2\sum_{\br,n} \left[ \frac{2S\Delta^z_\br(2h_\br)}{(i\nu_n)^2-(2h_\br)^2}\right ][b_{\br z}(n) b^*_{\br z}(n) + b_{\br z}^*(n) b_{\br z}(n)]
\end{eqnarray}
Note that, for convenience, we have defined $\chi_\br=-\cos\theta_\br\cos\phi_\br\sin\theta_\br$ and $\gamma_\br=2i\sin\theta_\br\sin\phi_\br.$ This leads us to the effective action and inverse Green function presented in the main text.

\section{High Temperature Phonon Spectral Function and pseudogap}

\begin{figure}
\includegraphics[width=0.8\textwidth]{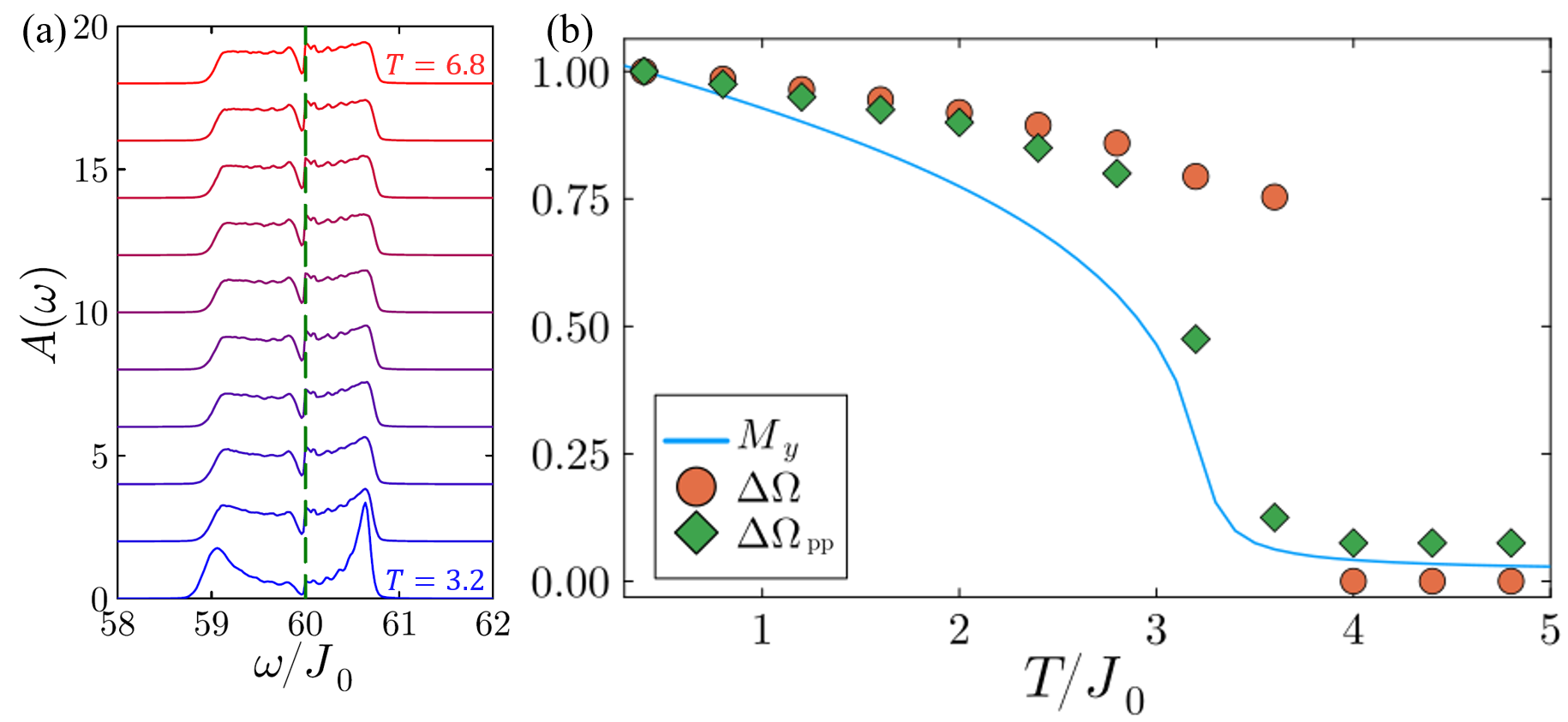}
\caption{{(a) Phonon spectral function $A(\omega)$ corresponding to the technique shown in Fig. 2 of the main text, for temperatures $T=3.2 J_0$ to $T=6.8J_0$ in the paramagnetic regime. Here, we more clearly observe the persistence of the pseudogap regime to $T^*\sim 1.2T_{\rm{oct}}$, as discussed in the main text. (vertical offset: 2) (b) Temperature dependence of the pseudo-chiral phonon splitting in the ferro-octupolar phase for both the phonon spectral function (orange circles) and the pump-probe spectral  function (green diamonds) absent phonon damping ($2\pi\eta/\Omega_0 =0$), plotted against the octupolar order parameter (solid blue line),  $M_y(T) = \frac{1}{N}
|\sum_{\br}\langle \tau_{\br y} \rangle|$. All curves are scaled to their lowest temperature value.}}
\label{fig:specFuncs-highT}
\end{figure}

In this section, we provide additional phonon spectral function data to accompany that shown in Fig. 2(a) of the main text. Interestingly, as discussed in the main text, when averaging $A(\omega)$ over sites in the lattice, we observe the presence of a pseudogap regime with increasing temperature. Explicitly, this refers to the presence of two broad, peaks at $\Omega_\pm \neq \Omega_0$ that persists to $T>T_{\rm{FO}}$, before the spectral weight fully broadens. We identify this regime where the edges of the single feature, produced by these two broad peaks, differ in spectral weight by \emph{more} than $10\%$ from the spectral weight at $\Omega_0$. We identify the pesudogap regime to persist to $T^*\simeq 1.2 T_{\rm{FO}}$, as shown in the phonon spectral functions in the range $T/J_0=(3.2,6.8)$ in Fig.\ref{fig:specFuncs-highT}(a). Further, we show the splitting of the undamped spectral functions for both the phonon ($\Delta\Omega$) and pump-probe spectral functions ($\Delta\Omega_{\rm pp}$) in Fig.\ref{fig:specFuncs-highT}(b). Note that, like Fig. 4 of the main text, neither curve is directly proportional to the order parameter, but both track the order parameter qualitatively. Here, we also note the presence of the pseudogap regime by the presence of finite splitting above $T_{\rm{FO}}$ but below $T^*$ in the phonon splitting $\Delta\Omega$.

\bibliography{main, ref2}